\documentclass[aps,prl,reprint]{revtex4-2}

\usepackage{amsmath, amssymb}
\usepackage{graphicx}

\begin{document}

\title{First-Order Trotter Error from a Second-Order Perspective}

\author{David Layden}
\email{david.layden@ibm.com}
\affiliation{IBM Quantum, Almaden Research Center, San Jose, California 95120, USA}

\begin{abstract}
Simulating quantum dynamics beyond the reach of classical computers is one of the main envisioned applications of quantum computers. The most promising quantum algorithms to this end in the near-term are the simplest, which use the Trotter formula and its higher-order variants to approximate the dynamics of interest. The approximation error of these algorithms is often poorly understood, even in the most basic cases, which are particularly relevant for experiments. Recent studies have reported anomalously low approximation error with unexpected scaling in such cases, which they attribute to quantum interference between the errors from different steps of the algorithm. Here we provide a simpler picture of these effects by relating the Trotter formula to its second-order variant. Our method generalizes state-of-the-art error bounds without the technical caveats of prior studies, and elucidates how each part of the total error arises from the underlying quantum circuit. We compare our bound to the true error numerically, and find a close match over many orders of magnitude in the simulation parameters. Our findings reduce the required circuit depth for the most basic quantum simulation algorithms, and illustrate a useful method for bounding simulation error more broadly.
\end{abstract}

\maketitle

One of the most promising applications of quantum computers is simulating the dynamics of other quantum systems that are beyond the reach of classical computers \cite{feynman:1982}. Notable examples are found in quantum chemistry \cite{aspuru-guzik:2005, babbush:2015, wecker:2014, poulin:2015, cao:2019}, quantum many-body physics \cite{raeisi:2012} and quantum field theory \cite{jordan:2012, jordan:2014}. A quantum simulation algorithm on an $n$-qubit quantum computer approximates evolution by a Hamiltonian $H$ for a time $t$ using quantum gates \footnote{We focus on the case where $H$ is time-independent.}. The first such algorithms relied on product formulas (PFs), which include the Trotter formula and its generalizations: if $H=\sum_{j=1}^L H_j$ where each $e^{-i H_j \delta t}$ can be exactly realized with quantum gates for any $\delta t$, then $e^{-iHt}$ can be approximated through a sequence of such unitaries by taking $\delta t$ as a sufficiently small timestep \cite{trotter:1959, lloyd:1996}. Distinct ``post-Trotter'' simulation algorithms not employing PFs have also been proposed in recent years, with better asymptotic performance \cite{berry:2014, berry:2015, low:2017, berry:2015b, low:2019, low:2018}. However, the relative experimental simplicity of PF algorithms and the small prefactors in their time complexity likely makes them---especially the simplest, low-order versions---the most relevant quantum simulation algorithms for near-term devices \cite{childs:2018, childs:2021}.

Lloyd gave the first PF algorithm, which relies on the first-order PF (PF1), or Trotter, formula \cite{lloyd:1996}. It breaks down the total time $t$ into $r$ timesteps of length $\delta t=t/r$ and approximates $e^{-iH \delta t}$ by $e^{-iH_L \delta t}\cdots e^{-iH_1 \delta t}$, introducing an $O(\delta t^2)$ error per timestep. (The PF order refers to the largest power of $\delta t$ to which the gates in a timestep match $e^{-iH\delta t}$.) The overall error from repeating this sequence $r$ times accumulates at most linearly with $r$ \footnote{$\|V_1 V_2 - W_1 W_2\| \le \|V_1-W_1\| + \|V_2-W_2\|$ for unitaries $V_i$ and $W_i$ and a unitarily-invariant norm $\|\cdot \|$.}, and therefore goes as $O(t^2/r)$, which can be made arbitrarily small (in principle) with large $r$. More precisely, implementing 
\begin{equation}
U_1 = \Big(
e^{-iH_L t/r}\cdots e^{-iH_1 t/r}
\Big)^r
\end{equation}
on an ideal quantum computer approximates $e^{-iHt}$ to a Trotter error of
\begin{equation}
\| e^{-iHt} - U_1 \| \le \frac{t^2}{2r}
\sum_{j=1}^L 
\Big \|
\sum_{k=j+1}^L
[H_k, H_j]
\Big \|,
\label{eq:PF1_error}
\end{equation}
where $\|\cdot\|$ is any sub-multiplicative, unitarily-invariant matrix norm, such as the operator/spectral, trace or Frobenius norms \cite{childs:2019, childs:2021}. This algorithm is readily generalized by visiting each $H_j$ multiple times per timestep according to a higher-order PF \cite{suzuki:1985, suzuki:1991}. For instance, the second-order PF (PF2) approximates $e^{-iHt}$ by
\begin{equation}
U_2 = 
\Big[
\Big(
e^{-iH_1 \frac{t}{2r}}\cdots e^{-iH_L \frac{t}{2r}}
\Big)
\Big(
e^{-iH_L \frac{t}{2r}}\cdots e^{-iH_1 \frac{t}{2r}}
\Big)
\Big]^r,
\end{equation}
which introduces an error of order $O(\delta t^3)$ per cycle and a cumulative error (which we will still call Trotter error for PF2 and higher-order PFs) of \cite{kivlichan:2020, childs:2021}
\begin{multline}
\| e^{-iHt} - U_2 \| \le \frac{t^3}{12 r^2}
\sum_{j=1}^L
\bigg( 
 \Big \|
\sum_{k, \ell=j+1}^L [H_\ell, [H_k, H_j]] 
\Big \| \\
+\frac{1}{2} \Big \|
\sum_{k=j+1}^L [H_j, [H_j, H_k]
\Big \| \bigg).
\label{eq:PF2_error}
\end{multline}
Several similar error bounds are also known. What many of these bounds have in common is their tendency to overestimate the true error, sometimes suggesting gate counts many orders of magnitude beyond what is empirically required for a given error tolerance \cite{childs:2018, babbush:2015, raeisi:2012}. In fact, numerical studies have reported Trotter errors that scale differently with $r$ and $t$ than most known bounds would suggest---sometimes displaying clearly different scalings in different regimes---even for the simplest instances of PF1 \cite{childs:2019, heyl:2019, tran:2020}.

Unexpectedly low Trotter error is a mixed blessing for quantum computing. With pre-fault-tolerant devices, one wishes to find a near-optimal balance between Trotter error and accumulated gate errors, which dominate at small and large $r$ respectively \cite{knee:2015, endo:2019, clinton:2020}. An overly loose bound on Trotter error makes this optimization difficult, thereby decreasing the overall accuracy when gate errors are taken into account. Likewise, a loose bound on Trotter error would make many simulations that are within reach for a given fault-tolerant quantum computer appear out of reach. This would be especially problematic for early fault-tolerant quantum computers, which are likely to be small. These issues highlight the pressing need for a better understanding of Trotter error, especially for low-order PF algorithms and Hamiltonians $H$ of particular relevance for near-term devices. 

Recent works by Heyl, Sieberer et al.\ \cite{heyl:2019, sieberer:2019} and Tran et al.\ \cite{tran:2020} have made significant advances to this effect. Heyl, Sieberer et al.\ \cite{heyl:2019, sieberer:2019} considered the transverse-field Ising model and showed, both numerically and perturbatively, that the error in simulating local observables with PF1 was much smaller than suggested by Eq.~\eqref{eq:PF1_error} and the like. Moreover, they found that for fixed timestep length $\delta t$ below some threshold value, the error was independent of the total simulation time $t$. The authors interpreted their findings through the lens of quantum chaos. Tran et al.\ \cite{tran:2020} generalized these results to the common setting where $H$ can be decomposed into $L=2$ realizable terms $H_1+H_2=H$, which encompasses all one-dimensional (1D) lattice Hamiltonians (i.e., spin chains) with nearest-neighbor interactions as well as some higher-dimensional models like the transverse-field Ising model with arbitrary interactions. They argued that the errors from different timesteps of PF1 interfere destructively and therefore accumulate sublinearly with $r$---an effect precluded in deriving Eq.~\eqref{eq:PF1_error}. For the special case of 1D nearest-neighbor lattice Hamiltonians, they bounded the Trotter error as $O(\frac{nt}{r} + \frac{nt^3}{r^2})$, whereas Eq.~\eqref{eq:PF1_error} predicts $O(\frac{n t^2}{r})$ scaling for this setting. The former scaling would seem optimal in $t$ for short times---even for post-Trotter methods \cite{berry:2015, haah:2021}---and, intriguingly, matches the $O(\frac{n t^3}{r^2})$ scaling of Eq.~\eqref{eq:PF2_error} for PF2 at long times. The authors conjectured that, for a fixed error tolerance, PF1 and PF2 have the same asymptotic gate counts for large $t$, but stopped short of proving this due to a technical limitation that introduced an $O(n^{1/2})$ overhead in their PF1 gate count compared to PF2.

\begin{figure}
    \centering
    \includegraphics[width=0.48\textwidth]{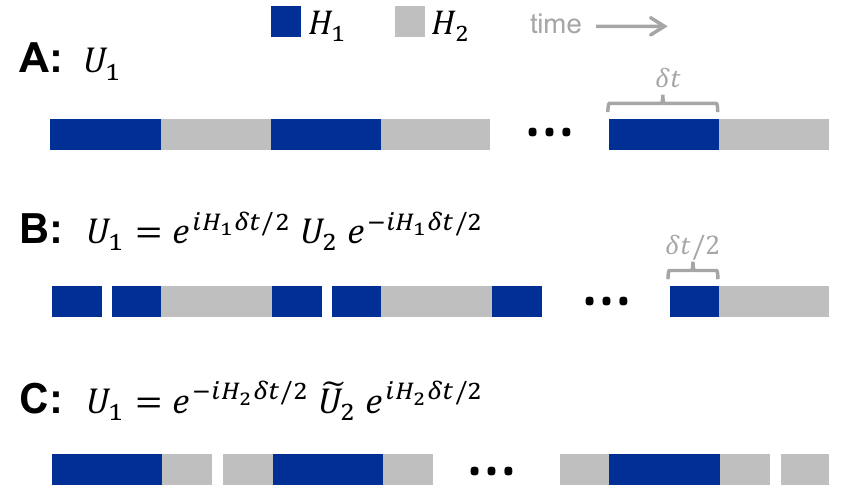}
    \caption{A graphical representation of how PF1 with $L=2$ \textit{(A)} can be understood as a sequence of PF2 timesteps plus additional layers at the beginning/end of the circuit \textit{(B)}. The same is true for PF2 with $H_1$ and $H_2$ exchanged \textit{(C)}.}
    \label{fig:sequence}
\end{figure}

\begin{figure*}
\includegraphics[width=\textwidth]{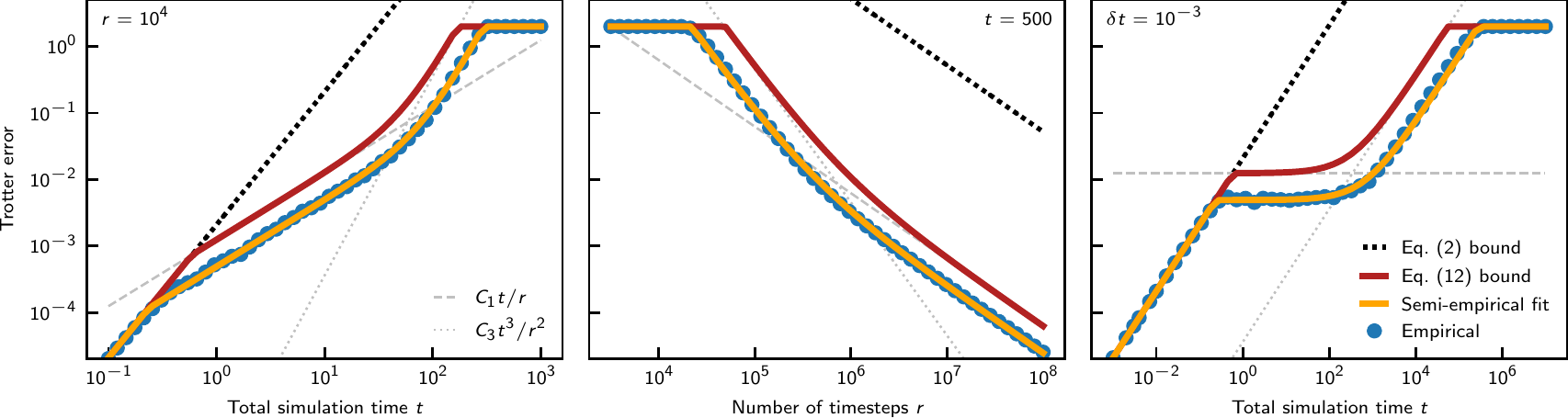}
\caption{The operator-norm PF1 error for a typical $n=10$ qubit instance of the 1D nearest-neighbor Heisenberg model with local disorder. The empirical Trotter error is compared with our bound from Eq.~\eqref{eq:main} and the earlier Eq.~\eqref{eq:PF1_error} bound of $C_2 t^2/r$, together with the other constituents of \eqref{eq:main}. The semi-empirical fit described in the text is also shown, using the same parameters $\alpha=2.53$ and $\beta=5.03$ in every panel. \textit{Left:} Trotter error versus the simulation time $t$ with the number of timesteps $r=10^4$ held fixed. \textit{Center:} Trotter error versus $r$ with $t=500$ held fixed. \textit{Right:} Trotter error versus $t$ with the length of each timestep $\delta t=t/r=10^{-3}$ held fixed. The values of $t$, $r$ and $\delta t$ in each panel were chosen to clearly illustrate the different behaviors predicted by Eq.~\eqref{eq:main} on a common scale. Equivalent plots for different $n$ and different disorder instances were qualitatively similar to those shown here.}
\label{fig:error}
\end{figure*}

Building on these recent works, we report an improved bound on PF1 Trotter error, and suggest a simple mechanism behind the aforementioned phenomena. In doing so, we prove the above conjecture of Tran et al.\ \cite{tran:2020}, find exact pre-factors, and offer an improved quantitative and qualitative agreement with numerics. We consider the setting introduced above, where $H$ can be decomposed for simulation into $L=2$ parts. ($H$ need not be a 1D nearest-neighbor model, though our setting encompasses this case.) The key observation is straightforward: when $L=2$ we can re-group the terms in $U_2$ as
\begin{equation}
\begin{aligned}
U_2 &= 
\Big( 
e^{-iH_1 \frac{t}{2r}} e^{-iH_2\frac{t}{r}}e^{-iH_1 \frac{t}{2r}} 
\Big)^r \\ 
&=
e^{-iH_1 \frac{t}{2r}} \,
\Big(
e^{-iH_2\frac{t}{r}}e^{-iH_1 \frac{t}{r}}
\Big)^r \,
e^{iH_1 \frac{t}{2r}} 
\end{aligned}
\end{equation}
since $e^{-iH_1 \frac{t}{2r}} \, e^{-iH_1 \frac{t}{2r}}$ can be compiled as $e^{-iH_1 \frac{t}{r}}$, as illustrated in Figs.~\ref{fig:sequence}A and \ref{fig:sequence}B. In other words, $U_1$ and $U_2$ differ only at the beginning and end of their quantum circuits:
\begin{equation}
U_1 = e^{iH_1 \frac{t}{2r}} \, U_2 \, e^{-iH_1 \frac{t}{2r}}.
\label{eq:U1_U2}
\end{equation}
From Eq.~\eqref{eq:U1_U2} we can quickly bound the PF1 Trotter error using the PF2 bound in Eq.~\eqref{eq:PF2_error} plus a term accounting for the beginning/end. Concretely:
\begin{equation}
\| e^{-iHt} - U_1 \| \le 
\|e^{-iHt} - U_2\| +
\big \| [e^{-iHt}, \, e^{-iH_1 \frac{t}{2r}} ] \big \|,
\end{equation}
where we substituted Eq.~\eqref{eq:U1_U2} for $U_1$, added and subtracted $e^{i H_1 \frac{t}{2r}} e^{-iHt} e^{-i H_1 \frac{t}{2r}}$, and used the triangle inequality and the norm's unitary invariance. The first term describes PF2 error and is bounded by Eq.~\eqref{eq:PF2_error}. For the second term, we use Kubo's identity
\begin{equation}
[A, e^{-i x B}] = -i e^{-i x B} \int_0^x e^{i s B} [A,B] e^{-i s B} ds
\label{eq:kubo}
\end{equation}
for matrices $A$ and $B$ \cite{suzuki:1985, kubo:1957}. Taking the norm of both sides for  $A=e^{-iHt}$ and $B=\frac{t}{2r}H_1$, then invoking unitary invariance and the triangle inequality to bring the norm into the integral gives
\begin{equation}
\big \| [e^{-iHt}, \, e^{-iH_1 \frac{t}{2r}} ] \big \|
\le 
\frac{t}{2r} \big \| [e^{-iHt}, H_1]  \big \|.
\label{eq:commutator_bound}
\end{equation}
We can similarly bound the right-hand side of \eqref{eq:commutator_bound} by $\frac{t}{r}\|H_1\|$, or by $\frac{t}{r}\|H_2\|$ since $[e^{-iHt}, H_1]=-[e^{-iHt}, H_2]$. Alternatively, we can apply Eq.~\eqref{eq:kubo} again to bound it by $\frac{t^2}{2r} \|[H_1,H_2]\|$, giving
\begin{equation}
\big \| [e^{-iHt}, \, e^{-iH_1 \frac{t}{2r}} ] \big \|
\le 
\frac{t}{r} \text{min} \Big( \|H_1 \|, \|H_2\|, \frac{t}{2} \big \|[H_1,H_2] \big \| 
\Big).
\end{equation}

To arrive at our main result, we note that PF1 is equally similar to PF2 with $H_1$ and $H_2$ exchanged, described by
\begin{equation}
\tilde{U}_2 = \Big( 
e^{-iH_2 \frac{t}{2r}} e^{-iH_1\frac{t}{r}}e^{-iH_2 \frac{t}{2r}} 
\Big)^r 
=
e^{i H_2 \frac{t}{2r}} \,  U_1 \,  e^{-i H_2 \frac{t}{2r}}
\end{equation}
and illustrated in Figs.~\ref{fig:sequence}A and \ref{fig:sequence}C. Repeating the steps above with $\tilde{U}_2$ instead of $U_2$ gives a similar bound on the Trotter error, but with $H_1$ and $H_2$ exchanged. Combining both bounds with the standard PF1 bound from \eqref{eq:PF1_error}, and noting that $\|V-W\|\le 2\|I \|$ for any two unitaries $V$ and $W$ gives our main result:
\begin{equation}
\| e^{-iHt} - U_1 \| 
\le \text{min}
\left(
\frac{C_2 t^2}{r}, \, \frac{C_1 t}{r} + \frac{C_3 t^3}{r^2}, \, 2 \| I \|
\right),
\label{eq:main}
\end{equation}
where 
\begin{equation}
\begin{gathered}
C_1 = \text{min} \big( \|H_1\|, \|H_2\| \big)
\qquad
C_2 = \frac{1}{2} \big \|[H_1,H_2] \big \| \\
C_3 = \frac{1}{12} \Big[ \text{min}(S) + \frac{1}{2} \text{max}(S) \Big]
\end{gathered}
\end{equation}
for
\begin{equation}
S = \Big\{
\big \| [H_1, [H_1, H_2]] \big \|, \, \big \| [H_2, [H_2, H_1]] \big \|
\Big \}.
\end{equation}
Note that this bound is symmetric under the exchange of $H_1$ and $H_2$ even though $U_1$ and $U_2$ are not, suggesting that it doesn't matter whether we start with $H_1$ or $H_2$; a feature known as ``ordering robustness'' \cite{childs:2019}.

In the special case where $H$ describes a 1D lattice of $n$ qubits with nearest-neighbor interactions, $C_j=O(n)$ for $j=1,2,3$ so Eq.~\eqref{eq:main} matches the bound of Tran et al.\ \cite{tran:2020} without additional caveats, and therefore proves their aforementioned conjecture \cite{tran:2020}.

We illustrate our bound by applying it to the nearest-neighbor Heisenberg model in 1D with random local disorder and open boundary conditions, for consistency with Refs.~\cite{childs:2018, childs:2019, tran:2020, childs:2021}. This model is described by $H=H_\text{even} + H_\text{odd}$ where
\begin{equation}
H_\text{even/odd} = \sum_{\substack{j=0\\j \text{ even/odd}}}^{n-2} \vec{\sigma}_j \cdot \vec{\sigma}_{j+1}
+ \sum_{\substack{j=0\\j \text{ even/odd}}}^{n-1} h_j Z_j, 
\end{equation}
$\vec{\sigma}_j = (X_j, Y_j, Z_j)$, and $h_j$'s are independent random variables describing disorder, each uniformly distributed over $[-1,1]$. The empirical PF1 error $\|e^{-iHt}-U_1\|$ (where $H_1 = H_\text{even}$, $H_2=H_\text{odd}$) for a typical instance is shown in Fig.~\ref{fig:error}, together with the earlier bound from Eq.~\eqref{eq:PF1_error} and our bound from Eq.~\eqref{eq:main}. Note the markedly improved quantitative and qualitative agreement of the latter. Fig.~\ref{fig:error} also shows a semi-empirical fit for the Trotter error, where we use the functional form of \eqref{eq:main} but replace the coefficients $C_1 \rightarrow C_1/\alpha$ and $C_3 \rightarrow C_3/\beta$, and fit $\alpha$ and $\beta$ to the empirical error. (We found no need to adjust $C_2$.) The resulting curves match the empirical error over many orders of magnitude with $\alpha$ and $\beta$ of order unity, displaying little variation with $n$ or between disorder instances. This close match suggests that the functional form of our bound may be optimal, but that the coefficients $C_1$ and $C_3$ could be somewhat improved.

Our result sheds new light on the mechanism behind the surprisingly complex PF1 error scaling observed in the most basic case of $L=2$, illustrated in Fig.~\ref{fig:error}. For instance, the near-equivalence between PF1 and PF2 in this case highlights the destructive interference between errors from different timesteps of PF1. The result suggests, however, that this interference happens primarily between adjacent timesteps---one can see this by (conceptually) slicing/regrouping PF1 timesteps into PF2 timesteps as in Fig.~\ref{fig:sequence}. Moreover, our simple derivation elucidates the source of each term in Eq.~\eqref{eq:main}. For instance, the discrepancy between PF1 and PF2 at the beginning/end of their circuits introduces the $O(t/r)$ term. When this term dominates, we get a regime in which the Trotter error for fixed $\delta t$ is nearly independent of $t$---visible in the right panel of Fig.~\ref{fig:error}. (This scaling with $t$ no longer seems meaningfully optimal when one considers its origin.) The $O(t^3/r^2)$ term, conversely, comes from the ``bulk'' (i.e., not the beginning/end) of the circuit, where PF1 and PF2 coincide. It can dominate for large $t$ or small $r$. Finally, the $O(t^2/r)$ term, which dominates at small $t$, arises simply when Eq.~\eqref{eq:PF1_error} gives the tightest error bound. The relative sizes of the terms in Eq.~\eqref{eq:main} define different scaling regimes with boundaries set by $C_1$, $C_2$ and $C_3$. 

Note that the error in simulating specific observables may be substantially lower than Eq.~\eqref{eq:main} would suggest. For instance, if the quantum computer is initialized in an eigenstate of $H_1$ ($H_2$) and measured in the eigenbasis of $H_1$ ($H_2$), then PF1 and PF2 become exactly equivalent and there will be no $O(t/r)$ term in the simulation error. This effect could occur in simulating the magnetization of the transverse-field Ising model initialized in a computational state, for example.

In summary, we considered the Trotter error in simulating a Hamiltonian $H$ using the first-order product formula when $H$ can be decomposed into $L=2$ realizable terms. Anomalously small error with unexpected scaling has recently been reported for this simple---albeit still quite general---setting, which is of particular interest for near-term quantum computers. We refined the tightest known bound on Trotter error for this setting through the straightforward observation that the first- and second-order product formulas are nearly identical when $L=2$. Crucially, our method clearly identifies the physical source of each constituent term in the Trotter error. Finally, we showed numerically that our bound can give a good quantitative and qualitative description of the true Trotter error, as well as a near-perfect fit over many orders of magnitude through a small adjustment of the constant prefactors.

Perhaps the simplest interpretation of our result is that one should likely use PF2 instead of PF1 for Hamiltonian simulation experiments when $L=2$. The resulting circuits are nearly identical, and the gates where they differ can contribute strongly to the PF1 Trotter error. More broadly, however, this work demonstrates an unconventional but useful way to think of Trotter error. Error bounds are typically found by bounding the error in a single timestep, then multiplying the result by the number of timesteps. However, we can just as well think of dividing the circuit up differently by grouping together gates from different timesteps. A judicious choice of partition can link product formulas of different orders. While our specific technique does not obviously generalize beyond PF1 with $L=2$, it could conceivably be adapted to find new error bounds or design new simulation algorithms in specific instances of interest beyond this setting. The numerous PF variants proposed in recent years may provide ample opportunity for such extensions \cite{hadfield:2018, campbell:2019, childs:2019b, ouyang:2020}. Given the prospects for simulating classically-intractable quantum dynamics on a quantum computer, the importance of accurately characterizing errors in such simulations can hardly be overstated.

\acknowledgements \textit{Acknowledgements.} We wish to thank Sergey Bravyi, Dmitri Maslov and Sarah Sheldon for helpful discussions.

\bibliography{references}

\end{document}